\documentclass{article}
\usepackage{graphicx} 
\usepackage{tikz}
\usepackage{hyperref}
\usetikzlibrary{positioning, fit, shapes.geometric}

\title{HandCT: hands-on computational dataset for X-Ray Computed Tomography and Machine-Learning}
\author{Valat Emilien, Valat Loth}

\begin{document}

\maketitle

\begin{abstract}
Machine-learning methods rely on sufficiently large dataset to learn data distributions. They are widely used in research in X-Ray Computed Tomography, from low-dose scan denoising to optimisation of the reconstruction process. The lack of datasets prevents the scalability of these methods to realistic 3D problems. We develop a 3D procedural dataset in order to produce samples for data-driven algorithms. It is made of a meshed model of a left hand and a script to randomly change its anatomic properties and pose whilst conserving realistic features. This open-source solution relies on the freeware Blender and its Python core. Blender handles the modelling, the mesh and the generation of the hand's pose, whilst Python processes file format conversion from obj file to matrix and functions to scale and center the volume for further processing. Dataset availability and quality drives research in machine-learning. We design a dataset that weighs few megabytes, provides truthful samples and proposes continuous enhancements using version control. We anticipate this work to be a starting point for anatomically accurate procedural datasets. For instance, by adding more internal features and fine tuning their X-Ray attenuation properties.

\end{abstract}

\section{Introduction}

\section*{Background \& Summary}
Machine-learning (ML) methods are widely used in X-Ray Computed Tomography (XCT), mainly for low-dose denoising, scarce-view CT and dual-energy CT. They rely on data samples to be trained on, and their performance is conditioned by the availability and quality of the datasets. At present, the limitations to the existing datasets we identify are related to the purposes for which they were designed as well as the difficulty to collect and use CT data.

\begin{figure}
    \centering
    \includegraphics{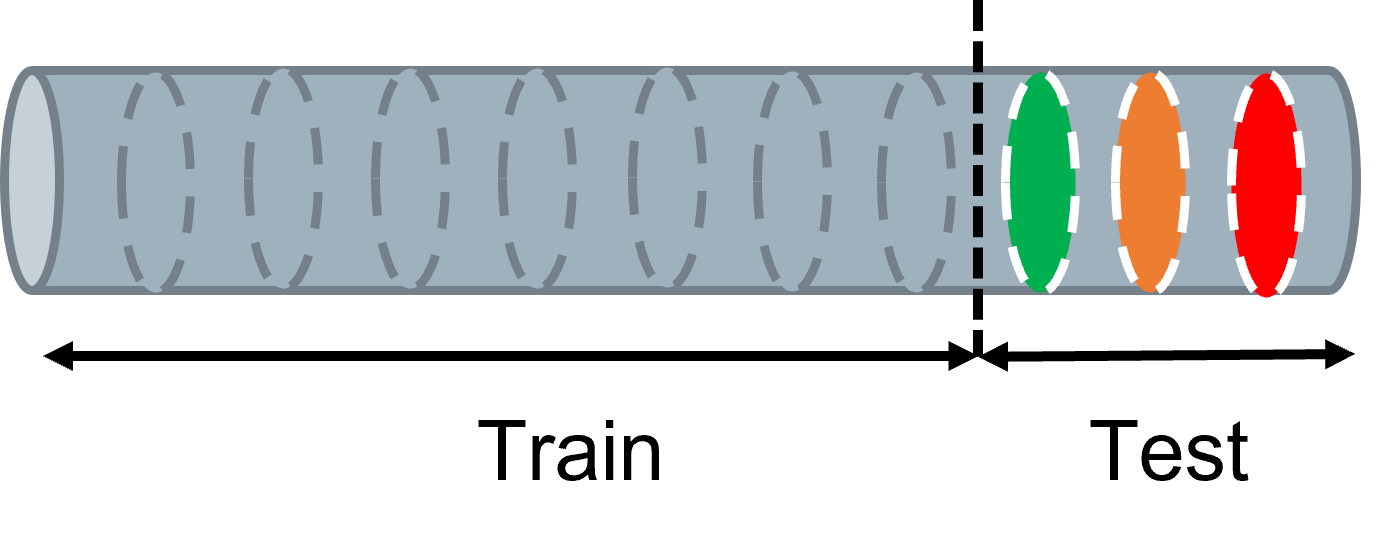}
    \caption{Example of data-leakage for 2D slices of a 3D volume. The color indicate the level of correlation with the last few training samples. Green, orange and red indicate strong, weak, poor correlation, respectively.}
    \label{fig:data_leakage}
\end{figure}

Benchmarking reconstruction algorithms and training ML approaches require different types of datasets. The Shepp-Logan phantom \cite{Shepp1974Fourier} is a computational phantom used routinely when comparing reconstruction processes. It consists of a set of ellipses with various X-Ray attenuation values and radius, and has been designed to resemble to a human brain. As training a ML algorithm on one data point is irrelevant, it cannot be used for that purpose. \cite{Hamalainen2015Tomographic} consists in a set of projections and an associated reconstruction of a single walnut, chosen for its likeness to a human brain. As all these items relate to the same data-point, it is unsuitable for ML purposes as well. The SophiaBeads dataset \cite{Coban2015Dataset} is used to benchmark reconstruction processes against a certain level of angular scarcity in the projection set. Even though there are different data-points in the angular domain, all the measures relate to one actual sample. \cite{Slotwinski2014Characterization, Carlo2018TomoBank, Singh2018Time} have the same limitation, knowing that the number of samples is to little to undertake ML procedures without over-fitting the data. One might consider using those datasets for 2D CT by reconstructing the 3D volumes, slice and re-project them to produce as many samples as there are slices. This can lead to data-leakage and should be avoided, as illustrated in Fig. \ref{fig:data_leakage}.

 When training image improvement methods on the first 180 slices out of 200 that compose a volume from the Sophiabeads dataset, the samples are highly correlated with their immediate neighbours. As such, when evaluating on the last 10\% of the dataset which corresponds to the last 20 slices, we observe that the performance of an approach depends on the index of the slice considered: the closer to the last slice of the training set a test sample is, the better. Averaging the performance across the test dataset might mask this, but is not very honest. Re-projecting data can be used in context where several volumes are present, as in \cite{Grove2015QuantitativeAdenocarcinoma, McCollough2016Overview}. 

CT datasets of real data are collected for and by CT practitioners. As underlined by \cite{Sarkissian2019Cone}, they rely on proprietary software and data formats not directly compatible with ML frameworks. For instance, \cite{Li2021MUG500} offers a collection of hundreds of Computer Assisted Design drawings of skulls, obtained by expert segmentation of CT scans of skulls. The use of this dataset requires considerable amount of space and is not shipped with file conversion scripts. The scientist coming from ML will often use unrealistic datasets made of ellipsoids, as in \cite{Jin2016Deep, Adler2017Learned, Kelly2017Deep}. The ease of use and development of these is out-weighted by the coarse feature modelled. To the extent of our knowledge, the only dataset that is designed for 3D Cone-Beam CT is \cite{Sarkissian2019Cone}. It is made of a considerable amount of scan data from 42 walnuts exactly fit for ML purposes. Each sample consists on average of 6 Gb of projection data. As such, there are few light-weight datasets of realistic computational phantoms for training machine-learning algorithms on. To the best of our knowledge, only \cite{Pelt2022Foam} offers a procedure to generate foam-like phantoms for X-Ray CT and machine-learning. To add to that literature, we develop HandCT, a procedural dataset that can generate sufficiently many good-quality samples for data-driven approaches.

It consists of a left-hand surface mesh modelled in Blender and a random morphological deformation routine in Python. Inspired by the Z-Anatomy project \cite{Kervyn2022First}, we chose to model an anatomically accurate object, that can be validated by anatomists and enhanced by 3D artists. To this initial layer, we added a Python script, implemented in Blender which offers a Python interface, that creates a sample by applying a modification (pose of the hand, pose of individual fingers, thickness of phalanxes) to the rig of the model, meshing it and saving it to an obj. format. We added a script to convert the latter to a Numpy matrix and to perform basic matrix operations (scaling and translating). The main feature of this dataset is the realistic anatomic feature modelling. The latter can be enhanced and refined by modifying the open-source mesh, and reproducibility is ensured using version control on the Zenodo and Github platforms. It is also important to note that the model itself weighs few megabytes.

\section*{Methods}
\begin{figure}[ht]
\centering
\begin{tikzpicture}[
    node distance = 7mm and -3mm,
every node/.style = {draw=black, rounded corners, fill=gray!30, 
                     minimum width=0cm, minimum height=0.5cm,
                     align=center}
                        ]

\draw node[fill=gray!10,rectangle, rounded corners ,minimum height=2cm, minimum width=4cm] at (1,0) {};     
\node[draw] at (-0.5,1) {Blender};

\draw node[fill=gray!10,rectangle, rounded corners ,minimum height=2cm, minimum width=2cm] at (5,0) {};     
\node[draw] at (5.5,1) {Python};

\node (base) at (0,0) {Base};
\node (sample) at (2,0) {Sample};
\node (array) at (5,0) {Array};

\draw [->,green] (base.north)to [out=30,in=150] (sample.north) ;
\draw [->,blue] (sample.south) to [out=-150,in=-30] (base.south);
\draw [->,red]  (sample.east) to (array.west);

\end{tikzpicture}
\caption{Data processing pipeline. The default model is first modified using random transformations (green arrow), then the morphed model is saved as a \textit{.obj} and the transformations reversed (blue arrow). The parsing operation (red arrow) to an array is irreversible as edge information is not stored.}
\label{fig:mother_child_array}

\end{figure}
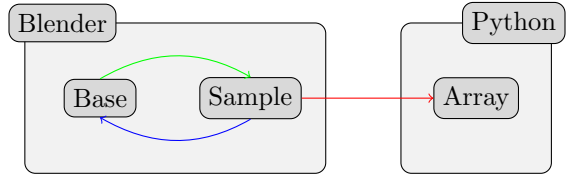

HandCT relies on three steps. First, a base model is designed and rigged by a 3D artist. Then, a python script modifies the latter into a sample model using random scaling of random parts of the rig. The sample's mesh is saved as an \textit{.obj} file and the modifications reversed back to the initial state. Finally, a python script converts the \textit{.obj} file into a numpy array and deals with absorption values, scaling of the volume in which the sample is defined and changing the coordinate system. The model is now saved as an array of desired size and orientation, ready to be used further. The interaction between base model, sample model and array is described in Figure \ref{fig:mother_child_array}.

\subsection*{Hand model design and rigging}
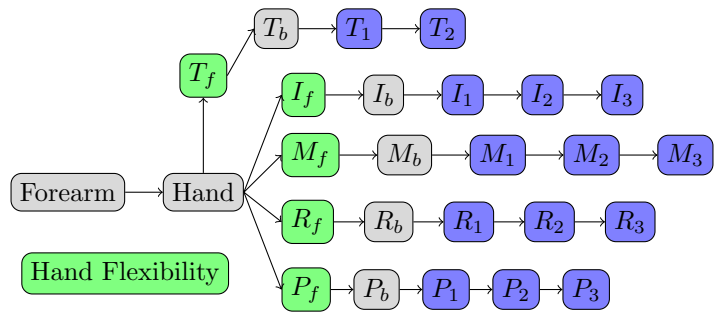
\begin{figure}
\centering
\begin{tikzpicture}[
    node distance = 7mm and -3mm,
every node/.style = {draw=black, fill=gray!30, rounded corners,
                        minimum width=0cm, minimum height=0.5cm,
                        align=center}
                        ]
\node (forearm) {Forearm};

\node (hand)[right=5mm of forearm.east] {Hand};

\node (hand_flexibility)[below=8mm of forearm.east, fill=green!50] {Hand Flexibility};

\node (mf)[above right=2mm and 5mm of hand.east, fill=green!50] {$M_f$};
\node (mb)[right=5mm of mf.east] {$M_b$};
\node (m1)[right=5mm of mb.east, fill=blue!50] {$M_1$};
\node (m2)[right=5mm of m1.east, fill=blue!50] {$M_2$};
\node (m3)[right=5mm of m2.east, fill=blue!50] {$M_3$};

\node (if)[above right=10mm and 5mm of hand.east, fill=green!50] {$I_f$};
\node (ib)[right=5mm of if.east] {$I_b$};
\node (i1)[right=5mm of ib.east, fill=blue!50] {$I_1$};
\node (i2)[right=5mm of i1.east, fill=blue!50] {$I_2$};
\node (i3)[right=5mm of i2.east, fill=blue!50] {$I_3$};

\node (rf)[below right=1mm and 5mm of hand.east, fill=green!50] {$R_f$};
\node (rb)[right=4mm of rf.east] {$R_b$};
\node (r1)[right=4mm of rb.east, fill=blue!50] {$R_1$};
\node (r2)[right=4mm of r1.east, fill=blue!50] {$R_2$};
\node (r3)[right=4mm of r2.east, fill=blue!50] {$R_3$};

\node (pf)[below right=10mm and 5mm of hand.east, fill=green!50] {$P_f$};
\node (pb)[right=3mm of pf.east] {$P_b$};
\node (p1)[right=3mm of pb.east, fill=blue!50] {$P_1$};
\node (p2)[right=3mm of p1.east, fill=blue!50] {$P_2$};
\node (p3)[right=3mm of p2.east, fill=blue!50] {$P_3$};

\node (tf)[above=10mm of hand.north, fill=green!50] {$T_f$};
\node (tb)[above right=5mm of tf.east] {$T_b$};
\node (t1)[right=5mm of tb.east, fill=blue!50] {$T_1$};
\node (t2)[right=5mm of t1.east, fill=blue!50] {$T_2$};

\draw [->,black] (forearm.east) to (hand.west);

\draw [->,black] (hand.north) to (tf.south);
\draw [->,black] (tf.east) to (tb.west);
\draw [->,black] (tb.east) to (t1.west);
\draw [->,black] (t1.east) to (t2.west);

\draw [->,black] (hand.east) to (if.west);
\draw [->,black] (if.east) to (ib.west);
\draw [->,black] (ib.east) to (i1.west);
\draw [->,black] (i1.east) to (i2.west);
\draw [->,black] (i2.east) to (i3.west);

\draw [->,black] (hand.east) to (mf.west);
\draw [->,black] (mf.east) to (mb.west);
\draw [->,black] (mb.east) to (m1.west);
\draw [->,black] (m1.east) to (m2.west);
\draw [->,black] (m2.east) to (m3.west);

\draw [->,black] (hand.east) to (rf.west);
\draw [->,black] (rf.east) to (rb.west);
\draw [->,black] (rb.east) to (r1.west);
\draw [->,black] (r1.east) to (r2.west);
\draw [->,black] (r2.east) to (r3.west);

\draw [->,black] (hand.east) to (pf.west);
\draw [->,black] (pf.east) to (pb.west);
\draw [->,black] (pb.east) to (p1.west);
\draw [->,black] (p1.east) to (p2.west);
\draw [->,black] (p2.east) to (p3.west);

\end{tikzpicture}

\caption{Hierarchical structure of the rig. Arrows indicate inheritance from parent to child. gray, green and blue nodes indicate rigid bones, pose-control and thickness-control nodes, respectively. Fingers are referred to with the first letter of their name.}
\label{fig:bone_tree}
\end{figure}

We build on a realistic 3D model initially designed for virtual and augmented reality applications \cite{Haupt2012Rigged}. This model is composed of two rigged hands (rig and mesh), but we kept only the left hand in this initial design. A rig is an underlying skeleton, not rendered at display time, that allow animators to give different poses to a model. Fig.\ref{fig:bone_tree} displays the skeleton hierarchy of our model. Compared to the original, we replaced the initial rigid bones by the so-called ``bendy bones'', allowing more finesse in the movements. There are six controllers for the hand pose: one per finger and an extra one for the whole hand at once. There are fourteen controllers for the phalanx thickness. The scale attribute of each controller is bounded between 0.75 and 1.15 using trial and error, to maintain a morphological coherence. 

\begin{figure}[ht]
    \centering
    \includegraphics[width=0.6\textwidth]{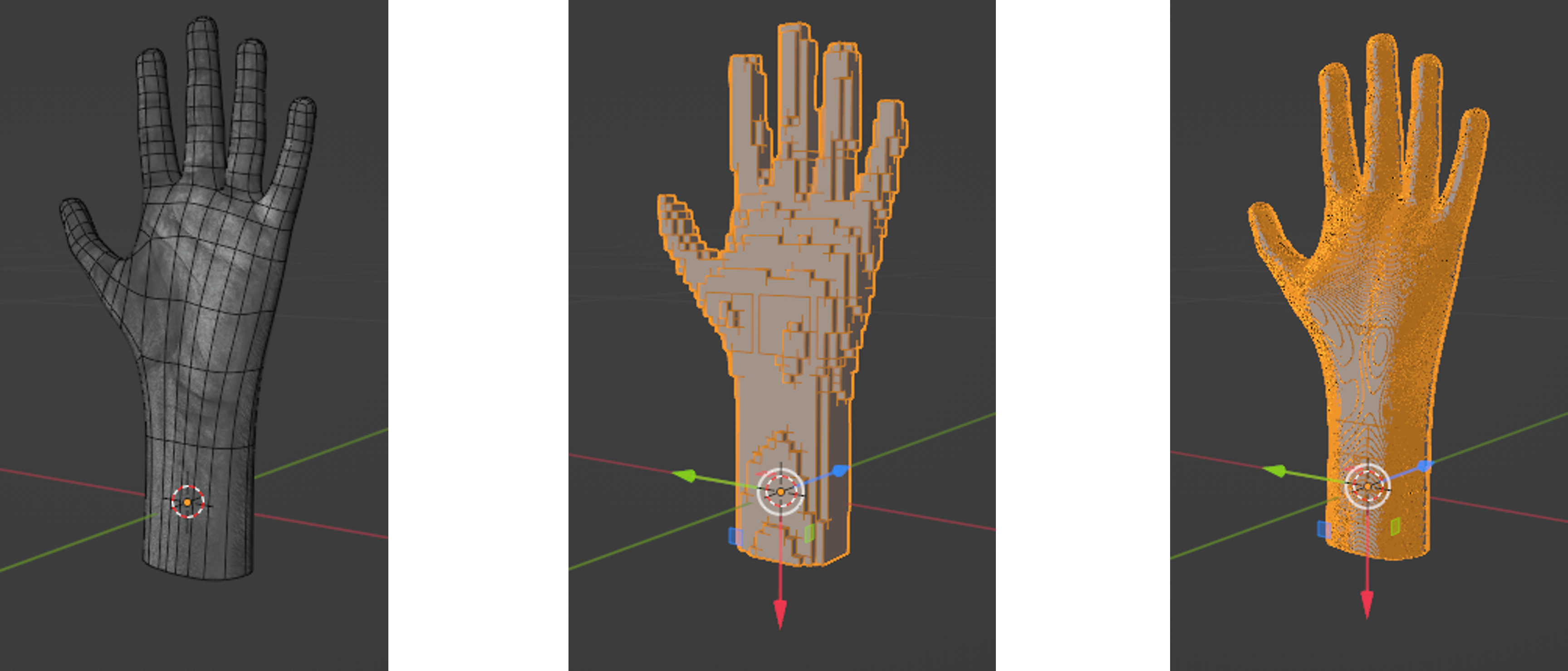}
    \caption{Comparison of the base model (Left), with a low-poly version (Middle) and the chosen resolution (Right).}
    \label{fig:octree_depth_comparison}
\end{figure}

The mesh controls the visual aspect of the hand. In order to have a sufficient resolution and visual smoothness, we used the ``Subsurf'' option to increase the density if vertices on the mesh. Then, in order to be processed, the model has to be remeshed to a high resolution point cloud. To perform this operation, we use the ``Block Remesh'' tool and set the Octree Depth parameter to a value of 10 using trial and error. Our way to control this parameter was to visually assess the quality of the model once loaded in Python, as shown in Fig. \ref{fig:octree_depth_comparison}. As a result, each model is composed of roughly 65k points. 

\subsection*{Python Processing Functions}
\begin{figure}
    \centering
    \includegraphics[width=0.7\textwidth]{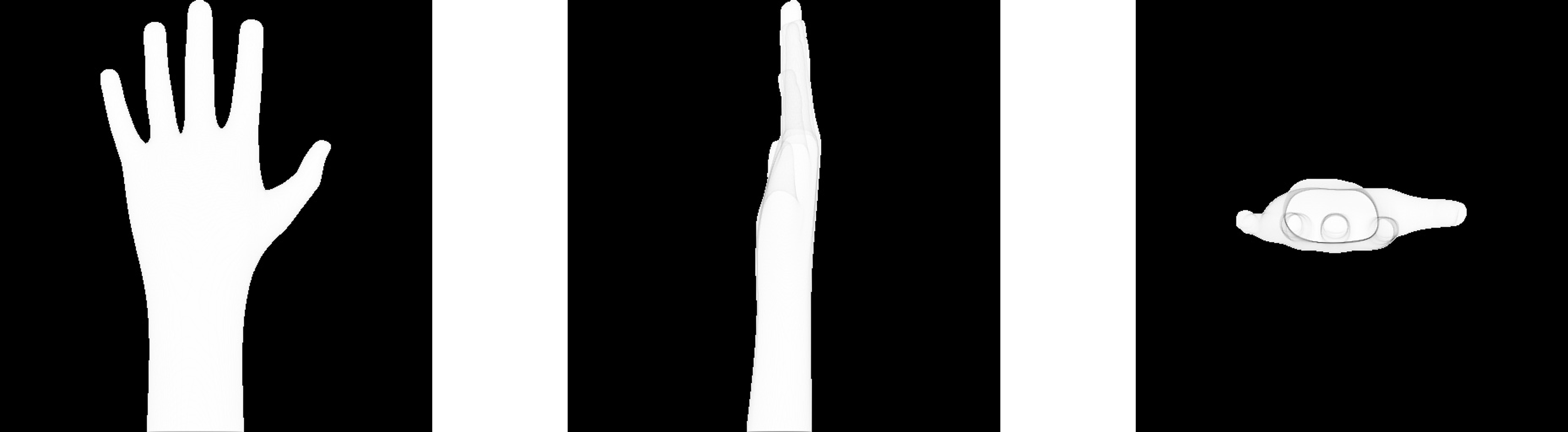}
    \caption{Comparison of the base model in front view (Left), side view (Middle) and bottom view (Right).}
    \label{fig:hand_views_array}
\end{figure}

We kept the Python-only part of the script at its minimum: the only functions that it handles are:
\begin{itemize}
    \item Randomly selecting a bendy bone and apply a random scaling.
    \item Parsing a .obj file to a point cloud.
    \item Scaling and centering the point cloud in a $512^3$ Numpy array.  
\end{itemize}
Fig. \ref{fig:hand_views_array} shows three views of the base model, once parsed to a converted to a Numpy matrix.

\section*{Data Records}
The data is stored on \href{https://zenodo.org/record/6473101}{Zenodo} and \href{https://github.com/Emvlt/procedural_hand_dataset}{GitHub}. The files are the following:

\begin{itemize}
    \item \texttt{base\textunderscore model.blend}. For the model and the morphological modifications script.
    \item \texttt{parse\textunderscore obj \textunderscore file.py}. For the centering, scaling and numpy conversion.
    \item \texttt{demo.py}. Can be run to display the base model once saved in Numpy format.
\end{itemize}

\section*{Technical Validation}
The purpose of this dataset is to give an easily modifiable yet accurate modelling of a hand. We bounded the range of deformations, as well as the some movements to preserve realism.

\section*{Usage Notes}
First, in Blender editor, edit the number of samples, the random seed used and the save path. Then, run the script \texttt{generate \textunderscore samples.py} from the Text Editor area. Running the script will generate and save the different samples, that can be further processed in Python using the command:
\smallbreak
\texttt{python obj\textunderscore to\textunderscore array.py ---path\textunderscore to\textunderscore file path\textunderscore to\textunderscore your\textunderscore file }
\smallbreak
Alternatively, the function \texttt{parse\textunderscore obj\textunderscore file} can be used in a loop iterating on the files path.

\section*{Code availability}

The dataset and the code are available on \href{https://zenodo.org/record/6473101}{Zenodo} and \href{https://github.com/Emvlt/procedural_hand_dataset}{GitHub}.

\bibliographystyle{abbrv}
\bibliography{references.bib}

\end{document}